\documentclass[twocolumn,prl,showpacs,amsfonts,amsmath,assymb,eufrak]{revtex4}

\usepackage{epsfig}
\usepackage{color}
\usepackage{bm}

\newcommand{\todayd}{\the\year/\the\month/\the\day}

\newcommand{\bib}{\bibitem}

\newcommand{\nui}{\mu}

\def \({\left(}
\def \){\right)}

\def\eq#1\en{\begin{equation}#1\end{equation}}  
\def\eqa#1\ena{\begin{align}#1\end{align}}
\def\eqg#1\eng{\begin{gather}#1\end{gather}}
\newcommand{\lb}[1]{\label{e:#1}}
\newcommand{\rlb}[1]{\eqref{e:#1}} 
\newcommand{\nl}{\notag\\}

\newcommand{\rbk}[1]{\left(#1\right)}

\newcommand{\sbkt}[1]{\langle#1\rangle}

\newcommand{\sumtwo}[2]%
{\mathop{\sum_{#1}}_{#2}}
\newcommand{\sumthree}[3]%
{\mathop{\mathop{\sum_{#1}}_{#2}}_{#3}}
\newcommand{\sumfour}[4]%
{\mathop{\mathop{\mathop{\sum_{#1}}_{#2}}_{#3}}_{#4}} 
\newcommand{\prodtwo}[2]%
{\mathop{\prod_{#1}}_{#2}}
\newcommand{\mintwo}[2]%
{\mathop{\min_{#1}}_{#2}}
\newcommand{\maxtwo}[2]%
{\mathop{\max_{#1}}_{#2}}
\newcommand{\maxthree}[3]%
{\mathop{\mathop{\max_{#1}}_{#2}}_{#3}}
\newcommand{\limtwo}[2]%
{\mathop{\lim_{#1}}_{#2}}
\newcommand{\suptwo}[2]%
{\mathop{\sup_{#1}}_{#2}}
\newcommand{\supthree}[3]%
{\mathop{\mathop{\sup_{#1}}_{#2}}_{#3}}
\newcommand{\supfour}[4]%
{\mathop{\mathop{\mathop{\sup_{#1}}_{#2}}_{#3}}_{#4}} 
\newcommand{\inftwo}[2]%
{\mathop{\inf_{#1}}_{#2}}
\newcommand{\infthree}[3]%
{\mathop{\mathop{\inf_{#1}}_{#2}}_{#3}}
\newcommand{\inffour}[4]%
{\mathop{\mathop{\mathop{\inf_{#1}}_{#2}}_{#3}}_{#4}} 

\newcommand\calC{{\cal C}}

\newcommand\calL{{\cal L}}

\newcommand\calP{{\cal P}}

\newcommand\calS{{\cal S}}

\newcommand{\bsr}{\boldsymbol{r}}

\newcommand{\bsv}{\boldsymbol{v}}

\newcommand{\bsF}{\boldsymbol{F}}

\newcommand{\bbR}{\mathbb{R}}
\newcommand{\bbZ}{\mathbb{Z}}
\newcommand{\ep}{\varepsilon}
\newcommand{\up}{\uparrow}

\newcommand{\Di}{\mathit{\Delta}}
\newcommand{\qedm}{\rule{1.5mm}{3mm}}
\newcommand{\fp}[2]{\dfrac{\partial #1}{\partial #2}}
\newcommand{\fpp}[2]{\dfrac{\partial^2 #1}{\partial{#2}^2}}
\newcommand{\fd}[2]{\dfrac{d#1}{d#2}}

\newcommand{\sumi}{\sum_{i=1}^N}

\newcommand{\sumnuo}{\sum_{\nu=1}^n}
\newcommand{\intt}{\int_0^\tau\hspace{-1mm}dt\,}
\newcommand{\DS}{\Di S}

\newcommand{\DK}{\Di K}
\newcommand{\la}{\lambda}
\newcommand{\hL}{\hat{\calL}}
\newcommand{\tA}{\tilde{A}}
\newcommand{\tR}{\tilde{R}}

\newcommand{\sigmat}{\sigma_{\rm tot}}
\newcommand{\bH}{\beta_{\rm H}}
\newcommand{\bL}{\beta_{\rm L}}
\newcommand{\QH}{Q_{\rm H}}
\newcommand{\QL}{Q_{\rm L}}
\newcommand{\etaC}{\eta_{\rm C}}

\newcommand{\para}[1]{{\em #1}\/.---}
\newcommand{\midskip}{\vspace{3pt}}

\begin{document}
\title{Universal trade-off relation between power and efficiency for heat engines}

\author{Naoto Shiraishi}
\affiliation{Department of Basic Science, The University of Tokyo, 
3-8-1 Komaba, Meguro-ku, Tokyo 153-8902, Japan}

\author{Keiji Saito}
\affiliation{Department of Physics, Keio University, 3-14-1 Hiyoshi, Yokohama 223-8522, Japan} 

\author{Hal Tasaki}
\affiliation{Department of Physics, Gakushuin University, 1-5-1 Mejiro, Toshima-ku, Tokyo 171-8588, Japan}

\date{\today}

\begin{abstract}
For a general thermodynamic system described as a Markov process, we prove a general lower bound for dissipation in terms of the square of the heat current, thus establishing that nonvanishing current inevitably implies dissipation. 
This leads to a universal trade-off relation between efficiency and power, with which we rigorously prove that a heat engine with nonvanishing power never attains the Carnot efficiency. Our theory applies to systems arbitrarily far from equilibrium, and does not assume any specific symmetry of the model.
\end{abstract}

\maketitle

Heat engines have been among central topics of thermodynamics since the seminal work of Carnot \cite{Carnot,callen}, who established that the efficiency of any heat engine operating with two heat baths cannot exceed the Carnot efficiency $\etaC$. 
In recent years considerable effort has been devoted to finding thermoelectric materials with higher efficiency \cite{mahan-rev,majundar-rev,dresselhaus,snyder}, and to fabricating stochastic cyclic heat engines in small systems \cite{steeneken2010,blickle2011,martinez2015,rosnagel2015,crivellari2014,koski2014}. 
It is crucial to develop fundamental understanding, on the basis of recent progress in nonequilibrium statistical mechanics \cite{seifert-rev}, about heat-to-work conversion mechanisms.

It is known, again since Carnot, that the Carnot efficiency can be achieved in quasi-static processes.
But the power, i.e., the work produced in a unit time, of  a quasi-static engine vanishes since it takes infinitely long time to complete a cycle.
Then a natural question arises whether there can be an engine with nonvanishing power which attains the Carnot efficiency.
This is indeed a special case of a fundamental question whether there is a universal trade-off relation between energy transfer and dissipation in thermodynamic processes.
Note that thermodynamics, which does not have the notion of time scale, cannot answer these questions.

There have been various attempts \cite{Benenti,Sothmann,Brandner,Balachandran,Brandner-full,Allahverdyan,Stark,Brandner-new,Sanchez,Holubec,Whitney,underdamped,Proesmans,Proesmans2,Sekimoto-Sasa, Aurell, RazSubasiPugatch2016, mintchev, Campisi, Ponmurugan} to look for engines with high efficiency and nonvanishing power. 
In particular Benenti, Saito, and Casati \cite{Benenti} studied the efficiency of thermoelectric transport in the linear response regime, and argued that broken time-reversal symmetry (caused e.g., by a magnetic field), that leads to nonsymmetrical Onsager matrix, might increase the efficiency; they even suggested that a cycle with nonvanishing power which operates reversibly may be realizable. 
At this level of argument, the restriction on the Onsager matrix elements imposed by the second law does not prohibit the coexistence of nonvanishing power with the Carnot efficiency.
This observation triggered a number of studies 
on the relation between power and efficiency \cite{Sothmann,Brandner, Balachandran, Brandner-full, Stark, underdamped, Proesmans, Proesmans2, Holubec, Whitney, Brandner-new,Sanchez,Allahverdyan,mintchev,Campisi, Ponmurugan, RazSubasiPugatch2016}.

Studies based on concrete models mainly within the linear response regime~\cite{Sothmann,Brandner,Balachandran,Brandner-full, Stark,Brandner-new,Sanchez,Holubec, Whitney, underdamped,Proesmans,Proesmans2,Sekimoto-Sasa, Aurell} have denied the possibility of engines with nonvanishing power and the maximum efficiency, suggesting a general no-go theorem.
See \cite{RazSubasiPugatch2016} where such a theorem for special models is obtained.
There still are a number of attempts, on the other hand, for the realization of  such engines~\cite{Allahverdyan, mintchev, Campisi, Ponmurugan}.
No matter what the current ``general belief" may be, it is desirable to have decisive conclusions on this fundamental issue without resorting to specific models, approximations, or restrictions (e.g., to the linear response regime).

In this Letter we present such general and rigorous results. 
We first prove a general lower bound for dissipation  (i.e., entropy production rate) in terms of the square of the total heat current to reservoirs.
The bound implies a universal trade-off relation between power and efficiency in heat engines, which, as a corollary, implies that a heat engine with nonvanishing power can never attain the Carnot efficiency.

Our theory applies to {\em any}\/ heat engine which is described by classical mechanics, and whose interaction with heat baths can be represented by a Markov process.
Practically speaking we cover essentially any realistic engines, macroscopic or mesoscopic, except those working in a genuine quantum regime.

Our trade-off relation relies essentially  only on the condition that the stochastic dynamics associated with a heat bath leaves the canonical distribution invariant.
We thus see that this condition is critical for the no-go theorem for an engine with nonvanishing power and the Carnot efficiency.

To get the present results, it was essential for us to look at this old 
problem in light of the notion of entropy production, which had been developed in the long and rich history of nonequilibrium statistical mechanics \cite{seifert-rev}.
In particular the idea of partial entropy production rate developed for Markov processes in \cite{SS,SIKS,SMS} played an important role.
Some of crucial ideas and techniques in the present Letter appeared in an unpublished article \cite{previous} by two of us (NS and KS).

\para{Main results}
Consider an arbitrary heat engine which undergoes a cyclic process with period $\tau$.
During a cycle, the engine may interact with $n$ external heat baths with finite inverse temperatures $\beta_1,\ldots,\beta_n$ in an arbitrary manner.
Let $J_\nu(t)$ be the heat current that flows from the engine to the $\nu$-th bath at time $t$.
The energy conservation implies that the total work done by the engine  is $W=-\sumnuo\intt J_\nu(t)$.
Define the total entropy production in the baths, which is a measure of dissipation in the cycle, by 
\eq
\DS:=\sumnuo\beta_\nu\intt J_\nu(t).
\lb{DSdef}
\en
It satisfies $\DS\ge0$, which is the second law.

Our main finding is the inequality
\eq
\Bigl(\intt\sumnuo|J_\nu(t)|\Bigr)^2\le\tau\,\bar{\Theta}\,\DS,
\lb{main0}
\en
which is proved for a general engine described by a Markov process.
Here $\bar{\Theta}$, which depends on the model and state, is always finite and proportional to the size of the engine \cite{en:size}.
For the  standard Langevin-type heat baths described by \rlb{Lnu}, one has $\bar{\Theta}=2\,\bar{\gamma}\bar{K}/\bar{\beta}\bar{m}$, where  $\bar{K}$ denotes the time average of the total kinetic energy of the engine, and $\bar{\beta}$, $\bar{\gamma}$, and $\bar{m}$ are properly averaged inverse temperatures (of the baths), the damping constant and the mass (of the engine), respectively.
See \rlb{Gt}.
Note that both the lhs and rhs of \rlb{main0} are proportional to the square of the size of the engine.
Therefore the inequality is meaningful in the thermodynamic limit as well.

The inequality \rlb{main0} manifests the fundamental trade-off relation: {\em nonvanishing current inevitably induces dissipation}\/.
To see the implication on efficiency of heat engines, consider the case with $n=2$ and let the inverse temperatures of the baths be $\bH$ and $\bL$ with $\bH<\bL$.
We denote, as usual, by $\QH>0$ the heat absorbed by the engine from the bath with $\bH$, and by $\QL>0$ the heat flowed from the engine to the bath with $\bL$.
The work is then $W=\QH-\QL$, and the entropy production is $\DS=\bL\QL-\bH\QH$.
The bound \rlb{main0} reduces to $(\QH+\QL)^2\le\tau\,\bar{\Theta}\,\DS$.

Let $\eta:=W/\QH$ be the efficiency of the engine, and $\etaC:=1-(\bH/\bL)$ be the Carnot efficiency.
Noting a relation in thermodynamics $\eta(\etaC-\eta)=W\DS/\{\bL(\QH)^2\}$~\cite{Whitney}, our bound yields a {\em trade-off relation between power and efficiency}
\eq
\frac{W}{\tau}\le\bar{\Theta}\,\bL\,\eta\,(\etaC-\eta).
\lb{PE}
\en
The averaged power $W/\tau$ must vanish as $\eta\up\etaC$ or (obviously) as $\eta\downarrow0$.
We conclude that {\em an engine with nonvanishing power never attains the maximum efficiency}\/.
The bound \rlb{PE} was discussed numerically in \cite{Brandner-new} for thermoelectric phenomena, and derived for Brownian heat engines with time reversal symmetry in \cite{underdamped}, both in the linear response regime.
It is proved here for systems arbitrarily far from equilibrium for general models without any specific symmetry.

\para{Setup and the main inequality}
Suppose that there are a heat engine, $n$ heat baths with inverse temperatures $\beta_1,\ldots,\beta_n$, and an external agent who operates on the engine (by, e.g., moving a piston, changing a potential, attaching or detaching heat baths).
Although our theorem applies to general Markov processes, we focus on a general classical engine modeled as a system of $N$ particles (with inertia) with arbitrary confining potential and interaction, possibly under magnetic field.
Let $m_i$, $\bsr_i$ and $\bsv_i$ denote the mass, the position and the velocity, respectively, of the $i$-th particle (with $i=1,\ldots,N$) \cite{en:ratchet}.
We collectively represent by $X=(\bsr_1,\ldots,\bsr_N;\bsv_1,\ldots,\bsv_N)$ the state of the system. 
We assume that the system is characterized by a set of parameters $\la$, which does not only determine the dynamics of the system (i.e., engine), but also the way it couples to the baths. 
We denote by $E^\la(X):=\sumi m_i|\bsv_i|^2/2+U^\la(\bsr_1,\ldots,\bsr_N)$ the total energy of the system with parameter $\la$.

The external agent varies the parameters according to a  fixed function $\la(t)$ of time $t$.
Let $\calP_t(X)$ be the probability density to find the system in $X$ at $t$.
It obeys the continuous master equation \cite{Vankampen,sekimoto}
\eq
\fp{}{t}\calP_t(X) = (\hL^{\la(t)}\calP_t)(X) \, . \lb{CM}
\en
The time evolution operator is decomposed into deterministic and dissipative parts as
\eq
\hL^\la = \hL^{0,\la} + \sum_{\nu=1}^n\sum_{i=1}^N \hL^{\nu,\la}_i.
\lb{Ldec}
\en
Here $\hL ^{0, \la}$ is the Liouville operator (see C of \cite{supp}) for the deterministic dynamics described by the Newton equation $m_i\ddot{\bsr}_i(t)=\bsF_i^{\la}(X)$. 
The force $\bsF_i^{\la}(X)$ consists of $-\boldsymbol{\nabla}_iU^\la(\bsr_1,\ldots,\bsr_N)$ and possible velocity dependent force (such as the Lorentz force).
The only assumption is that the resulting time evolution with fixed $\la$ preserves both the phase space volume and the total energy.

The operator $\hL^{\nu,\la}_i$ with $\nu=1,\ldots,n$ and $i=1,\ldots,N$ represents the dissipation of the $i$-th particle, i.e., the change in $\bsv_i$, caused by the $\nu$-th heat bath.
The most general expression reads \cite{Vankampen}
\eq
(\hL^{\nu,\la}_i \calP)(X):=
\int dY\{r^{\nu,\la}_i (X,Y)\calP(Y)-r^{\nu,\la}_i (Y,X)\calP(X)\},
\lb{Lnud}
\en
where $r^{\nu,\la}_i (X,Y)\ge0$ is the hopping rate from $Y$ to $X$. 
It leaves the canonical distribution with $\beta_\nu$ invariant, i.e., $\int dY\{r^{\nu,\la}_i(X,Y)e^{-\beta_\nu E^\la(Y)}-r^{\nu,\la}_i (Y,X)e^{-\beta_\nu E^\la(X)}\}=0$.
Discrete noise in  small engines such as the Rayleigh piston and the Brownian motor~\cite{Siegel, Meurs, Fruleux} can be represented by \rlb{Lnud} with suitably chosen $r^{\nu,\la}_i(X,Y)$, whose explicit form can be found, e.g., in eq.(2) of \cite{Meurs}.
In the limit where the change in velocity is infinitesimally small, \rlb{Lnud} reduces to
\eq
\hL^{\nu,\la}_i =
\frac{\gamma_\nu(\la,\bsr_i)}{m_i}\Bigl\{
\fp{}{\bsv_i}\cdot\bsv_i
+\frac{1}{\beta_\nu m_i}\fpp{}{\bsv_i}
\Bigr\},
\lb{Lnu}
\en 
which describes the standard Langevin noise \cite{Vankampen}.
With \rlb{Lnu}, the master equation \rlb{CM} becomes the Kramers equation. 
The ``damping constant'' $\gamma_\nu(\la,\bsr)$ represents the magnitude of noise from the $\nu$-th bath.
Note that it may depend on $\bsr$, and on $t$ through $\la(t)$.

We stress that the above formulation covers essentially any classical heat engines including the Brownian heat engine which was recently realized experimentally \cite{blickle2011,martinez2015} using a single particle in a harmonic trap \cite{en:td}. 
It is also easy to treat overdamped dynamics  \cite{SSTprep}.

The averaged heat current to the $\nu$-th bath at $t$ is defined in the standard manner (see A of \cite{supp}) as 
\eq
J_{\nu} (t):=-\sum_{i=1}^N\int dX E^{\la(t)}(X)(\hL^{\nu,\la(t)}_i \calP_t)(X).
\lb{Jnu}
\en
We then define the total entropy production rate in the system and the baths by
\eq
\sigmat(t):=\fd{}{t}H(\calP_t)+\sumnuo\beta_\nu J_{\nu} (t),
\lb{sigmat}
\en
where $H(\calP):=-\int dX\,\calP(X)\log\calP(X)$ is the Shannon entropy of the system.

The core of our theory is the inequality
\eq
\sumnuo |J_{\nu} (t)|\le\sqrt{\Theta(t)\,\sigmat(t)},
\lb{main1}
\en
which is valid for any $\calP_t$ satisfying the master equation \rlb{CM}. 
Here $\Theta(t)$ is a quantity which depends on the model and the state, but is finite and proportional to $N$.
For baths with \rlb{Lnu}, we have 
\eq
\Theta(t)=\sumi\sumnuo\frac{1}{\beta_\nu}\bigl\langle\gamma_{\nu} (\la(t),\bsr_i)\,
|\bsv_i|^2
\bigr\rangle_t, 
\lb{Gt}
\en 
where $\sbkt{\cdots}_t$ denotes the average with respect to $\calP_t$. 
See B of \cite{supp} for a concrete expression and an upper bound for $\Theta(t)$ for baths with \rlb{Lnud}.

To treat thermodynamic cycles of period $\tau$, we consider the case $\la(0)=\la(\tau)$, and assume $\calP_0=\calP_\tau$, which is always realized by running the cycle sufficiently many times.
We then define the total entropy production (in the baths) during a cycle by
\eq
\DS:=\intt\sigmat(t)=\intt\sumnuo \beta_\nu J_{\nu} (t),
\en
where the contribution from $H(\calP_t)$ vanished because of the cyclicity. 
It is essential that $\DS$ is written only in terms of the currents, which are measurable quantities.
By integrating \rlb{main1} over $t$, and using the Schwarz inequality, we readily obtain \rlb{main0}, whose implications have already been discussed, with $\bar{\Theta}:=\tau^{-1}\intt \Theta(t)$.

\para{Derivation}
We study the Markov jump process obtained by faithfully discretizing the continuous master equation \rlb{CM}.
We prove inequalities corresponding to \rlb{main1}, from which \rlb{main1} follows as continuum limits.
The mathematically minded reader should understand that we interpret  \rlb{CM} as a continuum limit of the master equation \rlb{ME}.

As usual we decompose the whole phase space into small $6N$-dimensional parallelepipeds whose size in the $v$-directions is $\ep$ and that in the $r$-directions is $\ep'$. Each cell is represented by $X$ at its center.

We now regard $X$ as a discrete variable, and denote by $E^\la_X$ the corresponding energy. The probability $p_{t,X}$ to find the system in $X$ at $t$ obeys the master equation
\eq
\fd{}{t}p_{t,X}=\sum_Y R^{\la(t)}_{XY}\,p_{t,Y},
\lb{ME}
\en
which is obtained as a discretization of \rlb{CM}. See C of \cite{supp} for the (standard) discretization procedure. 

As in \rlb{Ldec}, the transition rate is decomposed as $R^\la_{XY}= R^{0,\la}_{XY}+\sumnuo\sumi R^{\nu,i,\la}_{XY}$.
To simplify the notation we also write this as $R^\la_{XY}=\sum_{\nui}R^{\nui,\la}_{XY}$, where $\nui=0$ or $\nui=(\nu,i)$ with $\nu=1,\ldots,n$ and $i=1,\ldots,N$.
The transition rate for each $\nui$ satisfies $R^{\nui,\la}_{XY}\ge0$ for $X\ne Y$ and $\sum_XR^{\nui,\la}_{XY}=0$. For the deterministic part, we  assume that $\sum_YR^{0,\la}_{XY}=0$, which means that the uniform distribution is invariant under $R^{0,\la}_{XY}$. This property is always satisfied in the faithful discretization of a dynamics which preserves the phase space volume. 
For the dissipation of the $i$-th particle from the $\nu$-th bath, we assume the invariance of the corresponding canonical distribution, i.e., $\sum_YR^{\nu,i,\la}_{XY}e^{-\beta_\nu E^\la_Y}=0$. 

We decompose the heat current into contributions from each particle as $J_\nu(t)=\sumi J_{\nu,i}(t)$, where
\eq
J_{\nu,i} (t):=-\sum_{X,Y}E^{\la(t)}_XR^{\nu,i ,\la(t)}_{XY}p_{t,Y}=-\sum_{X,Y}K^{i}_X R^{\nu,i ,\la(t)}_{XY}p_{t,Y},  
\lb{Jnu2}
\en
with $K^{i}_X:=m_i |\bsv_i|^2/2$. 
We here noted that the dissipative dynamics changes only the velocity. 
We also decompose the change in the Shannon entropy $H(p):=-\sum_Xp_X\log p_X$ as 
\eq
\fd{}{t}H(p_t)=-\sum_X\dot{p}_{t,X}\log p_{t,X}=\sum_{\nui}\eta_{\nui} (t)
\en
with $\eta_{\nui}(t):=-\sum_{X,Y}R^{\mu,\la(t)}_{XY}p_{t,Y}\log p_{t,X}$. 
We then define the entropy production rate for $\nui$ by $\sigma_{\nui}(t):=\eta_{\nui}(t)+\beta_{\nui} J_{\nui}(t)$ with $\beta_0:=0$ and $\beta_{(\nu,i)}:=\beta_\nu$.
The total entropy production rate is written as $\sigmat(t)=\sum_{\nui} \sigma_{\nui}(t)$.

Define the dual transition rate \cite{en:dual} by $\tR^{\nui,\la}_{XY}:=e^{\beta_\nu(E^\la_Y-E^\la_X)}R^{\nui,\la}_{YX}$, which satisfies $\sum_X\tR^{\nui,\la}_{XY}=0$ because of the condition $\sum_YR^{\nui,\la}_{XY}e^{-\beta_\nu E^\la_Y}=0$. 
One then has
\eqa
\sigma_{\nui}(t)&=\sum_{X,Y}R^{\nui,\la(t)}_{XY}p_{t,Y}\log\frac{R^{\nui,\la(t)}_{XY}p_{t,Y}}{\tR^{\nui ,\la(t)}_{YX}p_{t,X}}
\nl
&=\sum_{X\ne Y}s(R^{\nui ,\la(t)}_{XY}p_{t,Y},\tR^{\nui ,\la(t)}_{YX}p_{t,X}),
\lb{sigmaii}
\ena
where the first expression is standard  \cite{seifert-rev} (see F of \cite{supp}), and the second with $s(a,b):=a\log(a/b)+b-a$ was introduced in \cite{SS,SIKS,SMS}, where the summand was named the partial entropy production rate. By using the inequality $s(a,b)\ge c_0(a-b)^2/(a+b)$ with $c_0=8/9$ (see E of \cite{supp}), and defining $\tA^{\nui,\pm}_{XY}:=R^{\nui,\la(t)}_{XY}p_{t,Y}\pm\tR^{\nui,\la(t)}_{YX}p_{t,X}$, we have 
\eq
\sigma_{\nui} (t)\ge c_0\sum_{X\ne Y}\frac{(\tA^{\nui ,-}_{XY})^2}{\tA^{\nui,+}_{XY}}.
\lb{sigmaAA}
\en

For $\nui =(\nu,i)$ we rewrite \rlb{Jnu2} as
\eq
J_\nui(t)=-\sum_{X\ne Y}\DK_X^i\tA^{\nui,-}_{X,Y}=-\sum_{X\ne Y}\DK_X^i\sqrt{\tA^{\nui,+}_{X,Y}}\,\frac{\tA^{\nui,-}_{X,Y}}{\sqrt{\tA^{\nui,+}_{X,Y}}},
\en
where $\DK_X^i:=K^{i }_X-\sbkt{K^{i}}_t$.
By using the Schwarz inequality and \rlb{sigmaAA}, and noting the relation $\sum_{Y(\ne X)}(\DK_X^i)^2\tR^{\nui,\la}_{YX}=\sum_{Y(\ne X)}(\DK_X^i)^2R^{\nui,\la}_{YX}$, which follows from $\tR^{\nui,\la}_{XX}=R^{\nui,\la}_{XX}$, we arrive at
\eq
|J_\nui (t)|\le\sqrt{\Theta^{(1)}_\nui (t)\,\sigma_\nui (t)}
\lb{main3}
\en
with 
\eq
\Theta^{(1)}_\nui (t):=\frac{1}{c_0}\sum_{X\ne Y}(\DK_X^i)^2
A^{\nui,+}_{XY} 
\lb{Gii}
\en
Here, we defined  $A^{\nui,\pm}_{XY}:=R^{\nui,\la(t)}_{XY}p_{t,Y}\pm R^{\nui,\la(t)}_{YX}p_{t,X}$.
By summing \rlb{main3} over $\nui$, applying the Schwarz inequality, and noting that \rlb{sigmaAA} implies $\sigma_0(t)\ge0$, 
we finally get 
$\sumnuo\sumi |J_{\nu , i} (t)|\le\sqrt{\Theta(t)\sigmat(t)}$ with  $\Theta(t)=\Theta^{(1)}(t):=\sumnuo\sumi \Theta^{(1)}_{\nu ,i} (t)$.
By taking the continuum limit, this implies the desired \rlb{main1}.
For discrete noise, where the rate $r^{\nu,\la}_i (X,Y)$ is finite, $\Theta^{(1)}(t)$ remains finite in the continuum limit (see B of \cite{supp}).

In the limit of Langevin noise with \rlb{Lnu} where $r^{\nu,\la}_i (X,Y)$ becomes singular, \rlb{main3} becomes meaningless since \rlb{Gii} diverges.
In this case we make use of the additional symmetry $R^{\nui,\la}_{XY}e^{-\beta_\nu E^\la_Y}=R^{\nui,\la}_{YX}e^{-\beta_\nu E^\la_X}$ (i.e., the detailed balance condition, see C of \cite{supp}) to derive a stronger bound with a new definition of $\Theta(t)$.
With the new symmetry, one easily verifies the standard expression \cite{seifert-rev} (see F of \cite{supp})
\eq
\sigma_\nui(t)=\sum_{X,Y}R^{\nui,\la(t)}_{XY}p_{t,Y}\log\frac{R^{\nui,\la(t)}_{XY}p_{t,Y}}{R^{\nui,\la(t)}_{YX}p_{t,X}},
\lb{sigmai}
\en
for $\mu\ne0$. 
By noting the symmetry between $X$ and $Y$ this can be written as
\eq
=\frac{1}{2}\sum_{X,Y}\{R^{\nui,\la(t)}_{XY}p_{t,Y}-R^{\nui,\la(t)}_{YX}p_{t,X}\}
\log\frac{R^{\nui,\la(t)}_{XY}p_{t,Y}}{R^{\nui,\la(t)}_{YX}p_{t,X}}.
\en
By using the inequality $(a-b)\log(a/b)\ge2(a-b)^2/(a+b)$ (see E of \cite{supp}), we find that
\eq
\sigma_\nui(t)\ge\sum_{X\ne Y}\frac{(A^{\nui,-}_{XY})^2}{A^{\nui,+}_{XY}}.
\lb{boundi}
\en
Again by using the symmetry, \rlb{Jnu2} is rewritten as
\eqa
J_\nui (t)&=-\sum_{X\ne Y}K^{i}_X A^{\nui,-}_{XY}
=-\frac{1}{2}\sum_{X\ne Y}(K^{i}_X-K^{i}_Y)A^{\nui,-}_{XY}
\nl
&=-\frac{1}{2}
\sum_{X\ne Y}(K^{i}_X-K^{i}_Y)\sqrt{A^{\nui,+}_{XY}}\,
\frac{A^{\nui,-}_{XY}}{\sqrt{A^{\nui,+}_{XY}}}, 
\lb{Jnu3}
\ena
which leads to \rlb{main3} with $\Theta^{(1)}_\nui (t)$ replaced by $\Theta^{(2)}_\nui (t)$:
\eqa
\Theta^{(2)}_\nui(t):=&\frac{1}{4}\sum_{X\ne Y}
(K^{i}_X-K^{i}_Y)^2A^{\nui,+}_{XY}
\nl
=&\frac{1}{2}\sum_{X\ne Y}
(K^{i}_X-K^{i}_Y)^2R^{\nui,\la(t)}_{XY}p_{t,Y}
\lb{Gi}
\ena
The continuum limit, which is now finite, is readily evaluated as in D of \cite{supp}, and we get \rlb{Gt} with $\Theta (t)=\Theta ^{(2)}(t):=\sumnuo\sumi \Theta^{(2)}_{\nu ,i} (t)$.

\para{Discussion}
We have proved that the power of a classical Markovian heat engine must vanish as its efficiency approaches the Carnot bound.
The essence was the trade-off relation \rlb{main0} which shows that any heat flux inevitably induces dissipation.
In ref.\cite{Ponmurugan}, attainability of nonvanishing power and the Carnot efficiency is discussed with the Onsager matrix in the classical regime. 
Our result denies the possibility of realizing this abstract proposal as a Markov process. 
The clarification of proposals based on quantum systems~\cite{Allahverdyan, mintchev, Campisi} is a next challenge. 
Toward this direction, extensions of the present results to the case where the engine exchanges quantum particles with particle baths will be discussed in~\cite{SSTprep}. 
Quantum cyclic heat engine will also be considered in~\cite{STSprep}.

From a theoretical point of view, the most basic result of ours is the inequality \rlb{main1}, which states for each moment that $\sigmat(t)$ is strictly positive whenever there is nonvanishing heat current. We must note that $\sigmat(t)$, which involves the change in the Shannon entropy, may not be a physically observable quantity. But if we are able to interpret $\sigmat(t)$ as a measure of instantaneous dissipation, the bound \rlb{main1} can be regarded as a more fundamental trade-off relation between heat current and dissipation, which is valid in any thermodynamic processes. See ~\cite{Maes} for a related observation.
It is interesting to apply the relation to transient processes.

When the state of the engine is close to equilibrium, the relation \rlb{main1} may be understood as follows. In order to have nonvanishing current $J$ between the engine and a bath, there should be a difference $\Di\beta$ in their inverse temperatures. Then the current $J$ induces the entropy production rate $\sigma\sim\Di\beta J$. Now if the current satisfies the linear response $J\simeq\kappa\Di\beta$, we have $\sigma\sim J^2/\kappa$.
The bound \rlb{main1}, which is $J^2\lesssim \Theta\sigma$, then reads $\kappa\lesssim \Theta$. Thus, at least everything is close to equilibrium, our trade-off relation boils down to an upper bound on the heat conductivity.
In fact, in close-to-equilibrium regime, we can show \cite{SSTprep} $\kappa\simeq\Theta^{(2)}$ for $\Theta^{(2)}$ of \rlb{Gt} or \rlb{Gi}.


\bigskip
It is a pleasure to thank Takashi Hara, whom we almost regard as a coauthor, for discussions and his essential contribution to the present work. We also thank 
Kay Brandner,
Tatsuhiko Koike,
Takashi Mori,
and
Hiroyasu Tajima
for useful discussions, and Shin-ichi Sasa and Yohei Nakayama for useful comments on the manuscript.
The present work was supported by Grant-in-Aid for JSPS Fellows No. 26-7602 (NS) and JSPS Grants-in-Aid for Scientific Research No. JP26400404 (KS) and No. JP16H02211 (KS and HT).




\clearpage

\makeatletter
\long\def\@makecaption#1#2{{
\advance\leftskip1cm
\advance\rightskip1cm
\vskip\abovecaptionskip
\sbox\@tempboxa{#1: #2}%
\ifdim \wd\@tempboxa >\hsize
 #1: #2\par
\else
\global \@minipagefalse
\hb@xt@\hsize{\hfil\box\@tempboxa\hfil}%
\fi
\vskip\belowcaptionskip}}
\makeatother
\newcommand{\vo}{\omega}

\begin{widetext}

\begin{center}
{\bf \Large Supplemental Material for ``Universal trade-off relation between power and efficiency for heat engines"}

\bigskip
Naoto Shiraishi, Keiji Saito, and Hal Tasaki
\end{center}

\bigskip

\begin{quotation}
Here we shall describe in detail some technical (or related) points which we did not discuss in the main text.
Although the main text is more or less self-contained and most of the topics here are standard, we hope that the reader may benefit if we collect them here.
\end{quotation}

\bigskip

\bigskip\noindent
{\bf \large A. Heat and work in a Markov process}
\midskip
\setcounter{equation}{0}
\def\theequation{A.\arabic{equation}}

Let us make some comments about the definition \rlb{Jnu} of the heat current for those readers not very familiar with approaches to nonequilibrium physics based on Markov processes.

Let the energy expectation value at time $t$ be
\eq
E(t):=\int dX\,E^{\la(t)}(X)\,\calP_t(X).
\lb{A:E}
\en 
Its time derivative is
\eq
\fd{}{t}E(t)=\int dX\,\dot{\la}(t)\rbk{\fd{E^{\la}(X)}{\la}}_{\la=\la(t)}\,\calP_t(X)
+\int dX\,E^{\la(t)}(X)\,\fp{}{t}\calP_t(X),
\lb{A:dE}
\en
where the two terms in the right-hand side are interpreted as contributions from mechanical work and from heat exchange, respectively, as follows.

The first term represents the change in the energy induced by the change of the functional form of $E^\la_X$, which is caused by the operation of the external agent.
One can imagine that the agent slightly changes the potential energy $U^\la(\bsr_1,\ldots,\bsr_N)$ of gas molecules, which corresponds, e.g., to moving a piston attached to a container.
We can thus regard the first term as a result of the exchange of energy between the system and the agent through mechanical means.
The second term then represents the change of energy caused by non-mechanical means; it should be identified with the heat transfer.

We thus define the power $P(t)$ to the external agent and the heat current $J(t)$ to the baths as
\eq
P(t)=-\int dX\,\dot{\la}(t)\rbk{\fd{E^{\la}(X)}{\la}}_{\la=\la(t)}\,\calP_t(X),
\lb{A:P}
\en
and
\eq
J(t)=-\int dX\,E^{\la(t)}(X)\,\fp{}{t}\calP_t(X),
\lb{A:J}
\en
respectively.

By definition we have $dE(t)/dt=-\{P(t)+J(t)\}$, which upon integration leads to the first law
\eq
W:=\intt P(t)=-\intt J(t)+E(0)-E(\tau).
\lb{A:W}
\en
For a cycle, where one has $E(0)=E(\tau)$, this means $W=-\intt J(t)$, the relation we discussed in the very beginning of the Letter.

By using the time-evolution equation \rlb{CM} and the decomposition $\hL^\la=\hL^{0,\la}+\sumnuo \sum_{i=1}^N \hL_i^{\nu,\la}$, the definition \rlb{A:J} of the current is rewritten as
\eq
J(t)=- \int dX\,E^{\la(t)}(X)(\hL^{0,\la(t)}\calP_t)(X) -\sumnuo \sum_{i=1}^N \int dX\,E^{\la(t)}(X)(\hL_i^{\nu,\la(t)}\calP_t)(X).
\lb{A:J2}
\en
We see that the contribution from $\nu=0$ is vanishing since $\hL^{0,\la}$ conserves the energy $E^\la(X)$.
We thus get 
\eq
J(t)=\sumnuo  J_{\nu}(t)
\en
with 
\eq
J_{\nu}(t):=-\sum_{i=1}^N\int dX\,E^{\la(t)}(X)(\hL_i^{\nu,\la(t)}\calP_t)(X),
\en
which recovers the definition \rlb{Jnu}.

\bigskip\noindent
{\bf \large B. Explicit form and upper bound of $\Theta^{(1)}(t)$ for discrete noise}
\midskip
\setcounter{equation}{0}
\def\theequation{B.\arabic{equation}}

Let us write down the explicit form of the function $\Theta^{(1)}(t)$ when the heat baths are described by \rlb{Lnud}.
As stated in the main text, we assume that the transition rate $r^{\nu,\la}_i (X,Y)\ge0$ is chosen so that the corresponding stochastic dynamics leaves the  canonical distribution invariant, i.e.,
\eq
\int dY\{r^{\nu,\la}_i (X,Y)e^{-\beta_\nu E^\la(Y)}-r^{\nu,\la}_i (Y,X)e^{-\beta_\nu E^\la(X)}\}=0,
\en
for any $\nu$, $i$, and $X$.

To get the desired expression for $\Theta^{(1)}(t)$ in the continuum limit of \rlb{Gii}, one simply substitutes the discretization \rlb{C:pP} and \rlb{C:Rr} and changes the sum in \rlb{Gii} into integral.
The result for the total $\Theta^{(1)}(t)$ then reads
\eq
\Theta^{(1)}(t)=\frac{1}{c_0}\sumnuo\sumi\int dX\,dY
\bigl\{K^{i}(X)-\sbkt{K^i}_t\bigr\}^2
\bigl\{r^{\nu,\la(t)}_i(X,Y)\,\calP_t(Y)+r^{\nu,\la(t)}_i(Y,X)\,\calP_t(X)\bigr\},
\lb{Thetanew}
\en
with $\sbkt{K^i}_t=\int dX\,K^i(X)\calP_t(X)$.

\bigskip

Let us derive a simple upper bound for $\Theta^{(1)}(t)$ of \rlb{Thetanew}.
This is important since it guarantees that $\Theta^{(1)}(t)$ is always bounded.

For simplicity we shall assume that the rate $r^{\nu,\la}_i (X,Y)$ satisfies the bound
\eq
\int dY\,r^{\nu,\la}_i(Y,X)\le A_\nu,
\lb{rbound}
\en
for any $\nu$, $i$, and $X$, where $A_\nu$ is a finite constant.
We note that the bound is satisfied for essentially all standard discrete noise.

Let us first note that \rlb{rbound} implies
\eq
\int dX\,dY
\bigl\{K^{i}(X)-\sbkt{K^i}_t\bigr\}^2
r^{\nu,\la(t)}_i(Y,X)\,\calP_t(X)
\le
A_\nu\int dX
\bigl\{K^{i}(X)-\sbkt{K^i}_t\bigr\}^2\calP_t(X)
=A_\nu\Bigl\langle\bigl\{K^{i}-\sbkt{K^i}_t\bigr\}^2\Bigr\rangle_t.
\lb{KrA}
\en
Note that $\sbkt{\{K^{i}-\sbkt{K^i}_t\}^2}_t$ is (the square of) the fluctuation of kinetic energy, which should be finite in any physically meaningful state.

To treat the remaining term, we observe that
\eqa
\int dX\,dY
&\bigl\{K^{i}(X)-\sbkt{K^i}_t\bigr\}^2
r^{\nu,\la(t)}_i(X,Y)\,\calP_t(Y)
=
\int dX\,dY
\bigl\{K^{i}(X)-K^{i}(Y)+K^{i}(Y)-\sbkt{K^i}_t\bigr\}^2
r^{\nu,\la(t)}_i(X,Y)\,\calP_t(Y)
\nl
&\le2\int dX\,dY\bigl\{K^{i}(X)-K^{i}(Y)\bigr\}^2
r^{\nu,\la(t)}_i(X,Y)\,\calP_t(Y)
+2\int dX\,dY\bigl\{K^{i}(Y)-\sbkt{K^i}_t\bigr\}^2
r^{\nu,\la(t)}_i(X,Y)\,\calP_t(Y),\nonumber
\intertext{
where we used the inequality $(a+b)^2\le2(a^2+b^2)$.
As in \rlb{KrA} the above is further bounded as
}
&\le 2\sbkt{(\delta K^{i,\nu})^2}_t+2A_\nu\Bigl\langle\bigl\{K^{i}-\sbkt{K^i}_t\bigr\}^2\Bigr\rangle_t.
\lb{KrA2}
\ena
where $\sbkt{(\delta K^{i,\nu})^2}_t:=\int dX\,dY\bigl\{K^{i}(X)-K^{i}(Y)\bigr\}^2
r^{\nu,\la(t)}_i(X,Y)\,\calP_t(Y)$ is the expectation value of the square of the jump in the kinetic energy of the $i$-th paritlce.
This is also expected to be finite.

By summing up the contributions from \rlb{KrA} and \rlb{KrA2}, we finally get
\eq
\Theta^{(1)}(t)\le
\frac{1}{c_0}\sumnuo\sumi
\Bigl\{2\sbkt{(\delta K^{i,\nu})^2}_t+3A_\nu\Bigl\langle\bigl\{K^{i}-\sbkt{K^i}_t\bigr\}^2\Bigr\rangle_t\Bigr\},
\en
which guarantees the important fact that $\Theta^{(1)}(t)$ always remains finite and is at most proportional to $N$.

\bigskip
{\em Remark:}
The rate corresponding to the Langevin type noise described by \rlb{Lnu} does {\em not}\/ satisfy the bound \rlb{rbound}, as is clear from the expression \rlb{C:RS} below.
This is of course not a problem since we have a much better expression \rlb{Gt} of $\Theta(t)$ from which its finiteness is obvious.

\bigskip\noindent
{\bf \large C. Procedure of discretization}
\midskip
\setcounter{equation}{0}
\def\theequation{C.\arabic{equation}}

Although the discretization procedure we use may be rather standard, we shall explain it here for completeness.

The original phase space is the Euclidean space $\bbR^{6N}$ (or its subspace), whose element is $X=(\bsr_1,\ldots,\bsr_N;\bsv_1,\ldots,\bsv_N)$.
We decompose the phase space into the union of a small $6N$-dimensional parallelepiped whose size in the $v$-directions is $\ep$ and that in the $r$-directions is $\ep'$.
We denote by $\vo:=(\ep'\ep)^{3N}$ the volume of a cell.
We shall represent each cell by $X$ at its center.
See Figure~\ref{f:1}.

\begin{figure}
\centerline{\epsfig{file=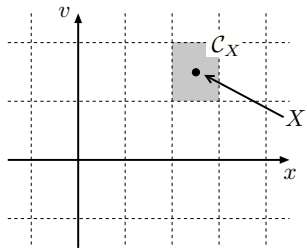,width=4cm}}
\caption[dummy]{
The phase space $\bbR^2$ is decomposed into cells with size $\ep \times \ep'$.
A cell is represented by the coordinate $X$ at its center, which is here indicated by the black dot.
The cell including $X$ is denoted as $\calC_X$, which is indicated by the gray region.
}
\label{f:1}
\end{figure}

Let $\Lambda$ be the collection of all $X$ at the center of a cell.
$\Lambda$ can be identified with the $6N$-dimensional lattice $(\ep'\bbZ)^{3N}\times(\ep\bbZ)^{3N}$.
For each $X\in\Lambda$, we denote by $\calC_X\subset\bbR^{6N}$ the cell  centered at $X$.

Let $X\in\Lambda$.
The probability $p_{t,X}$ for the discrete model is related to the probability density $\calP_t(X)$ of the continuum model by
\eq
p_{t,X}\simeq\int_{X'\in\calC_X}dX'\,\calP_t(X')\simeq\vo\,\calP_t(X).
\lb{C:pP}
\en
We shall design the discrete master equation \rlb{ME} so that it, with the identification \rlb{C:pP}, converges to the continuous master equation \rlb{CM}.

\bigskip
\noindent
{\bf Deterministic part:}~Let us start with the deterministic part defined by $\hL^{0,\la}$. 
The Liouville operator is given by
\eq
\hL^{0,\la}:=\sumi
\Bigl\{
-\bsv_i\cdot\fp{}{\bsr_i}-
\frac{1}{m_i}\fp{}{\bsv_i}\cdot\bsF_i^{\la}(X)
\Bigr\} ,
\lb{L0}
\en 
where $\bsF_i^{\la}(X)$ is the force acting on the $i$-th particle.

The procedure for determining the corresponding transition rate $R_{YX}^{0,\la}$ is as follows.

We first prepare the uniform distribution on the cell $\calC_X$.
Then each point in the phase space evolves according to the Newton equation with the force $\bsF_i^\la(X)$ for a short time $\Di t$, where we keep the parameters $\la$ fixed.
See Figure~\ref{f:2}.
Let $\operatorname{Prob}(\Di t,Y)$ be the probability to find the state in the cell $\calC_Y$ after the time evolution.
The desired transition rate for $Y\ne X$ is determined by
\eq
R^{0,\la}_{YX}:=\lim_{\Di t\downarrow0}
\frac{1}{\Di t}\operatorname{Prob}(\Di t,Y).
\lb{C:R0}
\en
Then the diagonal element $R^{0,\la}_{XX}$ is determined so that $\sum_{Y}R^{0,\la}_{YX}=0$ holds.

From this construction it is obvious that the transition rate $R^{0,\la}_{YX}$ leaves the uniform distribution (in a properly defined finite subset of the phase space) invariant, and hence $\sum_XR^{0,\la}_{YX}=0$.

\begin{figure}
\centerline{\epsfig{file=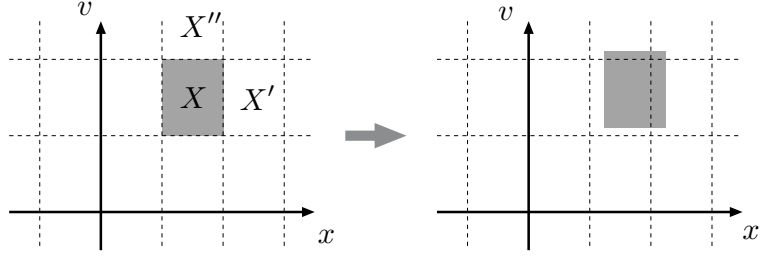,width=10cm}}
\caption[dummy]{
The procedure for determining the transition rate $R^{0,\la}_{YX}$ corresponding to the deterministic time-evolution governed by $\hL^{0,\la}$. We start from the uniform distribution in the cell $\calC_X$, and let the states evolve deterministically for time $\Di t$. }
\label{f:2}
\end{figure}

\bigskip
Let us check, for completeness, that this procedure really recovers the desired \rlb{L0}.
For simplicity we shall examine the simplest case where the system has only one degree of freedom, and hence $X=(x,v)\in\bbR^2$.
Extension to higher degrees of freedom is automatic (although formulas may become complicated).

Let the force be $F(x,v)$.
Then the time evolution for $\Di t$ is given by
\eq
x\ \to\ x+v\Di t+O((\Di t)^2),\quad 
v\ \to\ v+\frac{F(x,v)}{m}\Di t+O((\Di t)^2).
\en
Let $X=(x,v)\in\Lambda$ label the initial cell, and consider two neighboring cells labeled by $X'=(x+\ep',v)\in\Lambda$ and $X''=(x,v+\ep)\in\Lambda$.
See Figure~\ref{f:2}.
The probability of finding the state in these nearby cells are calculated (to the lowest order in $\ep$ and $\ep'$) as
\eqa
\operatorname{Prob}(\Di t,X')
=\frac{1}{\vo}\ep\,v\,\Di t+O((\Di t)^2)
=&\frac{1}{\ep'}v\,\Di t+O((\Di t)^2),
\\
\operatorname{Prob}(\Di t,X'')
=\frac{1}{\vo}\ep'\frac{F(x,v)}{m}\Di t+O((\Di t)^2)
=&\frac{1}{\ep}\frac{F(x,v)}{m}\Di t+O((\Di t)^2).
\ena
We thus find
\eq
R^{0}_{Y,X}=
\begin{cases}
\dfrac{1}{\ep'}v&\text{if $Y=X'$}\\
\dfrac{1}{\ep}\dfrac{F(x,v)}{m}&\text{if $Y=X''$}\\
0&\text{if $Y\not\in\{X,X',X''\}$}.
\end{cases}
\lb{C:R0YX}
\en
The diagonal element is given by
\eq
R^0_{XX}=-\sumtwo{Y\in\Lambda}{(Y\ne X)}R^0_{YX}=-\Bigl(\dfrac{1}{\ep'}v+\dfrac{1}{\ep}\dfrac{F(x,v)}{m}\Bigr).
\lb{C:R0XX}
\en

With these transition rates, the master equation \rlb{ME} becomes
\eqa
\fd{}{t}p_{t,(x,v)}
&=R^0_{(x,v),(x-\ep',v)}\,p_{t,(x-\ep',v)}+R^0_{(x,v),(x,v-\ep)}\,p_{t,(x,v-\ep)}
+R^0_{(x,v),(x,v)}\,p_{t,(x,v)}
\nl&=\dfrac{1}{\ep'}v\,p_{t,(x-\ep',v)}+
\dfrac{1}{\ep}\dfrac{F(x,v-\ep)}{m}\,p_{t,(x,v-\ep)}
-\Bigl(\dfrac{1}{\ep'}v+\dfrac{1}{\ep}\dfrac{F(x,v)}{m}\Bigr)\,p_{t,(x,v)}
\nl
&=v\frac{1}{\ep'}(p_{t,(x-\ep',v)}-p_{t,(x,v)})
+\frac{1}{m\ep}\Bigl(F(x,v-\ep)p_{t,(x,v-\ep)}-F(x,v)p_{t,(x,v)}\Bigr) ,
\ena
where $(x,v),(x-\ep',v),(x,v-\ep)\in\Lambda$.
This is approximated as 
\eq
\fd{}{t}p_{t,(x,v)}\simeq-v\fp{}{x}p_{t,(x,v)}-\frac{1}{m}\fp{}{v}\bigl(F(x,v)p_{t,(x,v)}\bigr),
\en 
which clearly recovers \rlb{L0}.

Obviously the energy is not exactly conserved after this discretization.
But the violation becomes smaller as $\ep$ and $\ep'$ approach zero, and the conservation is recovered in the continuum limit.

Similarly, although the deterministic time-evolution in terms of $\hL^{0,\la(t)}$ conserves the Shannon entropy (even when $\la(t)$ depends on time), the evolution by $R^{0,\la(t)}_{XY}$ does not.
Again the violation is small when the distribution is broad (compared with the size of cells), and one recovers the conservation in the continuum limit.

In the main text, we did {\em not} show that the entropy production from $R^{0,\la(t)}_{XY}$ vanishes in the continuum limit, but only proved that $\sigma_0(t)\ge0$ in general.
This inequality is sufficient for our purpose of proving the main inequality.

\bigskip
\noindent
{\bf Kramers type heat bath:}~We next discuss the discretization $R^{\nu \, i,\la}_{XY}$ of the time-evolution operator \rlb{Lnu}, which describes a Kramers type heat bath.
This part may be the least trivial.

The rate $R^{\nu,i,\la}_{XY}$ with $X\ne Y$ is nonvanishing only when the cells $X,Y$ are neighboring in a $v$-direction, or more precisely, $X$ and $Y$ differ by $\ep$ only in one of the components of $\bsv_i$.
Let us denote the relevant components of $X$ and $Y$ as $v_X$ and $v_Y$, respectively.
They satisfy $|v_X-v_Y|=\ep$.
Then we set 
\eq
R^{\nu,i,\la}_{XY}=\frac{\gamma(\la,\bsr_i)}{m_i^2\beta_\nu\ep^2}
e^{(\beta_\nu m_i/4)(v_Y^2-v_X^2)}.
\lb{C:RS}
\en 
Noting that
\eq
\frac{m_i}{2}(v_Y^2-v_X^2)=E^\la_Y-E^\la_X,
\en
one finds that the rate satisfies the detailed balance condition
\eq
R^{\nu,i,\la}_{XY}\,e^{-\beta_\nu E^\la_Y}
=R^{\nu,i,\la}_{YX}\,e^{-\beta_\nu E^\la_X}.
\lb{C:DB}
\en
The diagonal elements $R^{\nu,i,\la}_{XX}$ are again determined from the condition $\sum_YR^{\nu,i,\la}_{YX}=0$.

\bigskip

We must verify that this transition rate \rlb{C:RS} recovers the Kramers-type heat bath described by \rlb{Lnu}.
We shall again, for simplicity, treat the case with a single degree of freedom, i.e., $X=(x,v)$.
Since the variable $x$ does not change under the transition rate \rlb{C:RS}, we simply omit $x$, and regard $v$ (which is an integer multiple of $\ep$) as the label of cells.
We also consider a single heat bath, and denote $\beta_\nu$ as $\beta$, and also write $\gamma_\nu(\la,\bsr)$ as $\gamma$.

Let  $A=\gamma/(m^2\beta\ep^2)$.
The transition rate \rlb{C:RS} is
\eq
R_{v,v'}=A\,e^{(\beta m/4)(-v^2+{v'}^2)},
\en
for $|v-v'|=\ep$.

We now evaluate relevant rates by expanding in $\ep$ to get
\eqg
R_{v\pm\ep,v}=A\,e^{(\beta m/4)(\mp 2v\ep-\ep^2)}
=A\Bigl\{1\mp\frac{\beta m v}{2}\ep+\frac{(\beta mv)^2}{8}\ep^2-\frac{\beta m}{4}\ep^2+O(\ep^3)\Bigr\},\\
R_{v,v\pm\ep}=A\,e^{(\beta m/4)(\pm 2v\ep+\ep^2)}
=A\Bigl\{1\pm\frac{\beta m v}{2}\ep+\frac{(\beta mv)^2}{8}\ep^2+\frac{\beta m}{4}\ep^2+O(\ep^3)\Bigr\}.
\eng

With these transition rates, the master equation \rlb{ME} becomes
\eqa
\fd{}{t}p_{t,v}&=
R_{v,v+\ep}\,p_{t,v+\ep}+R_{v,v-\ep}\,p_{t,v-\ep}
-(R_{v+\ep,v}+R_{v-\ep,v})\,p_{t,v}
\nl
&=A\Bigl\{1+\frac{(\beta mv)^2}{8}\ep^2\Bigr\}
(p_{t,v+\ep}+p_{t,v-\ep}-2p_{t,v})
+A\frac{\beta m v}{2}\ep\,(p_{t,v+\ep}-p_{t,v-\ep})
\nl&\hspace{0.3cm}
+A\frac{\beta m}{4}\ep^2(p_{t,v+\ep}+p_{t,v-\ep}+2p_{t,v})+O(\ep).
\ena
By substituting $A$, and letting $\ep\downarrow0$, this reduces to 
\eq
\fd{}{t}p_{t,v}=\frac{\gamma}{m}\fp{}{v}(v\,p_{t,v})+\frac{\gamma}{m^2\beta}\fpp{}{v}p_{t,v},
\en 
which precisely recovers \rlb{Lnu}.

\bigskip
\noindent
{\bf Discrete noise:}~As for baths with discrete noise described by \rlb{Lnud}, discretization is trivial.

For $X,Y\in\Lambda$ with $X\ne Y$, we set
\eq
R^{\nu,i,\la}_{XY}=\vo\,r^{\nu,\la}_i(X,Y),
\lb{C:Rr}
\en
and determine $R^{\nu,i,\la}_{XX}$ from $\sum_YR^{\nu,i,\la}_{YX}=0$.

\bigskip
\noindent
{\bf Definitions of entropy:}~The correspondence between the continuum and the discrete descriptions is trivial for most quantities, but one needs to be slightly careful about the Shannon entropy.
Let $p_X$ be a discrete probability, and $\calP(X)$ be the corresponding probability density, i.e., $p_X\simeq\vo\calP(X)$.
Then note that
\eq
H_{\rm disc}(p)=-\sum_Xp_X\log p_X\simeq-\sum_X\vo\calP(X)\log\bigl\{\vo\calP(X)\bigr\}
\simeq-\int dX\,\calP(X)\log\calP(X)-\vo\log\vo=H_{\rm cont}(\calP)-\vo\log\vo,
\en
which means that the discrete entropy $H_{\rm disc}(p)$ and the continuum entropy $H_{\rm cont}(\calP)$ differ by an ``infinite constant" $\vo\log\vo$.
But this discrepancy does not cause any problems, since we are always interested in the change in the entropy.

\bigskip\noindent
{\bf \large D. Continuum limit}
\midskip
\setcounter{equation}{0}
\def\theequation{D.\arabic{equation}}

Let us make a short comment about the derivation of the neat formula \rlb{Gt} for engines coupled to Kramers-type heat baths.

We start from the expression \rlb{Gi} of $\Theta^{(2)}_\nu(t)$, and use the fact that the transitions rates have specific form \rlb{C:RS}.

Fix $Y\in\Lambda$, and suppose $R^{\nu i,\la}_{XY}\ne0$.
Then only one component of $\bsv_i$ for some $i$ is different between $X$ and $Y$.
Denoting the relevant component of $Y$ as $v$, we see that
\eq
(K^i_X-K^i_Y)^2=\Bigl(\frac{m_i}{2}(v\pm\ep)^2-\frac{m_i}{2}v^2\Bigr)^2
=(m_iv\ep)^2+O(\ep^3).
\en
Since the corresponding transition rate is 
\eq
R^{\nu, i , \la}_{XY}=\frac{\gamma(\la,\bsr_i)}{m_i^2\beta_\nu\ep^2}
\,e^{O(\ep)},
\en 
we find 
\eq
\frac{1}{2}\sumtwo{X\in\Lambda}{(X\ne Y)}
(K^{i}_X-K^{i}_Y)^2 R^{\nu, i,\la}_{XY}
=\frac{1}{2}
m_i^2\,2|\bsv_i|^2\ep^2
\frac{\gamma(\la,\bsr_i)}{m_i^2\beta_\nu\ep^2}
+O(\ep)
= \gamma(\la,\bsr_i)\frac{|\bsv_i|^2}{\beta_\nu}+O(\ep).
\en 
The desired \rlb{Gt} then follows.

\bigskip\noindent
{\bf \large E. Proof of the inequalities}
\midskip
\setcounter{equation}{0}
\def\theequation{E.\arabic{equation}}

Let us prove two elementary inequalities used in the ``Derivation''.
We learned the inequalities and their proof from Takashi Hara.

\bigskip
The first inequality is
\eq
(a-b)\log\frac{a}{b}\ge\frac{2(a-b)^2}{a+b},
\lb{E:Hara1}
\en
for any $a,b>0$.

\bigskip\noindent
{\em Proof}\/:
Because of the symmetry we can assume $a>b$.
We will then prove that
\eq
\log\frac{a}{b}\ge\frac{2(a-b)}{a+b}.
\en
Note that
\eq
\log\frac{a}{b}-\frac{2(a-b)}{a+b}=-\log u-\frac{2(1-u)}{1+u}=:g(u),
\en
where $u:=b/a$ satisfies $0<u<1$.
Because $g(1)=0$, and
\eq
g'(u)=-\frac{(1-u)^2}{u(1+u)^2}\le0,
\en
we see $g(u)\ge0$ for $0<u<1$.~\qedm

\bigskip

The second inequality is
\eq
a\log\frac{a}{b}+b-a\ge c_0\frac{(a-b)^2}{a+b},
\lb{E:Hara2}
\en
for any $a,b>0$,
where the constant $c_0$ satisfies $c_0\le8/9$. (See below for an improvement.)

\bigskip\noindent
{\em Proof}\/:
Let the constant $c_0$ be such that $0<c_0\le1$, and note that
\eq
\frac{1}{a}\Bigl(a\log\frac{a}{b}+b-a-\frac{c_0(a-b)^2}{a+b}\Bigr)
=\log\frac{a}{b}+\frac{b}{a}-1-\frac{c_0(1-b/a)^2}{1+b/a}
=-\log u+u-1-\frac{c_0(1-u)^2}{1+u}=:h(u),
\en
where $u:=b/a>0$.
We shall prove $h(u)\ge0$ for $c_0\le8/9$.

We have $h(1)=0$, and the first derivative is
\eq
h'(u)=\frac{u-1}{u(1+u)^2}\bigl\{(1-c_0)u^2+(2-3c_0)u+1\bigr\}.
\lb{hprime}
\en

Suppose first that $0<u<1$.
Then we have
\eq
\frac{u-1}{u(1+u)^2}<0,
\en
and
\eq
(1-c_0)u^2+(2-3c)u+1\ge0-u+1\ge-1+1=0,
\en
where we noted that $c_0\le1$ implies $2-3c_0\ge-1$.
We thus see $h'(u)\le0$, and conclude that $h(u)\ge0$ if $0<u\le1$.

Next suppose that $u>1$, where we have 
\eq
\frac{u-1}{u(1+u)^2}>0.
\en
Also observe that
\eq
(1-c_0)u^2+(2-3c_0)u+1=(1-c_0)\Bigl\{u+\frac{2-3c_0}{2(1-c_0)}\Bigr\}^2
+\frac{c_0(8-9c_0)}{4(1-c_0)}\ge\frac{c_0(8-9c_0)}{4(1-c_0)},
\en
where the right-hand side is nonnegative if $c_0\le8/9$.
In this case we have $h'(u)\ge0$, and hence $h(u)\ge0$.~\qedm

\bigskip

In fact, the condition $c_0\le8/9$ for the constant $c_0$ is not optimal.
Let us show how to get the optimal constant.

From \rlb{hprime}, one finds that the equation $h'(u)=0$ (for fixed $c_0$) has at most three solutions, and the largest solution is given by
\eq
u^*(c_0)=\frac{3c_0-2+\sqrt{9(c_0)^2-8c_0}}{2(1-c_0)}.
\en
It is easy to check that $h(u)\geq 0$ holds for all $u$ if and only if $h(u^*(c_0))\geq 0$.
Let $c^*$ denote $c_0$ such that $h(u^*(c_0))=0$.
We find that the inequality \rlb{E:Hara2} is valid for $c_0\le c^*$.

The equation $h(u^*(c_0))=0$ can be solved numerically, and we find the optimal constant to be $c^*=0.89612\cdots$.

\bigskip\noindent
{\bf \large F. On relative entropy}
\midskip
\setcounter{equation}{0}
\def\theequation{F.\arabic{equation}}

In our theory the strictly positive lower bounds \rlb{boundi} and \rlb{sigmaAA} of the entropy production rates were essential.
Weaker result, namely, the nonnegativity of entropy production rate is well-known, and can be proved easily \cite{sekimoto}.
In the present note, we shall review the standard information theoretic proof of the nonnegativity, which makes use of relative entropy (or the Kullback-Leibler divergence).
The proof is not only of interest, but also sheds light on the peculiar  expressions \rlb{sigmai} and \rlb{sigmaii} of the entropy production rates, which also played important roles in our theory.

For background, see, e.g,
T. M. Cover and J. A. Thomas,
{\em Elements of Information Theory}\/,
(Wiley-Interscience, 2006).
(But our presentation is not exactly the same as the one found in this book.)

Denote the states as $x,y,\ldots\in\calS$, and let $(R_{xy})_{x,y\in\calS}$ be an arbitrary transition rate matrix; it satisfies $R_{xy}\ge0$ if $x\ne y$ and $\sum_{x\in\calS}R_{xy}=0$.
Let $(q_x)_{x\in\calS}$ be the corresponding stationary distribution, i.e., $\sum_{y\in\calS}R_{xy}q_y=0$.
We also assume $q_x>0$ for any $x\in\calS$.
Then for any probability distribution $(p_x)_{x\in\calS}$, one has
\eq
\sum_{x,y\in\calS}R_{xy}p_y\,\log\frac{q_x}{p_x}\ge0.
\lb{F:1}
\en

Let us remark that this inequality leads to the following very general ``H-theorem''.
Let $p_{t,x}$ obey the master equation
\eq
\fd{}{t}p_{t,x}=\sum_{y\in\calS}R_{xy}\,p_{t,y}.
\en
Define the relative entropy by $D(p_t|q):=\sum_{x\in\calS}p_{t,x}\log(p_{t,x}/q_x)$.
Then one has the monotonicity
\eq
\fd{}{t}D(p_t|q)=\sum_{x\in\calS}R_{xy}p_{t,y}\log\frac{p_{t,x}}{q_x}\le0.
\en

\bigskip\noindent{\em Proof of \rlb{F:1}}\/:
We first define the dual transition rate by 
\eq
\tR_{yx}:=\frac{R_{xy}q_y}{q_x}.
\en
One easily verifies that $\tR_{yx}\ge0$ if $y\ne x$, and $\sum_{y\in\calS}\tR_{yx}=0$.
It also holds that $\sum_{x\in\calS}\tR_{yx}q_x=0$, i.e., $\tR$ also has $(q_x)_{x\in\calS}$ as its stationary distribution.
Then note that
\eq
\sum_{x,y\in\calS}R_{xy}p_y\,\log\frac{q_x}{p_x}=
\sum_{x,y\in\calS}R_{xy}p_y\,\log\frac{R_{xy}q_x}{R_{xy}p_x}
=
\sum_{x,y\in\calS}R_{xy}p_y\,\log\frac{R_{xy}q_xp_y}{R_{xy}p_xq_y}
=
\sumtwo{x,y\in\calS}{(x\ne y)}R_{xy}p_y\,\log\frac{R_{xy}p_y}{\tR_{yx}p_x}.
\lb{F:RRRR}
\en
Since $R_{xx}p_x=\tR_{xx}p_x$, we see that
\eq
\sumtwo{x,y\in\calS}{(x\ne y)}R_{xy}p_y=
\sumtwo{x,y\in\calS}{(x\ne y)}\tR_{yx}p_x=:C.
\en
For $x\ne y$, define
\eq
P_{(x,y)}:=\frac{R_{xy}p_y}{C},\quad Q_{(x,y)}:=\frac{\tR_{yx}p_x}{C},
\en
which can be viewed as probability distributions for the states $(x,y)$ with $x,y\in\calS$ and $x\ne y$.
Then
\eq
\sum_{x,y\in\calS}R_{xy}p_y\,\log\frac{q_x}{p_x}=
C\sumtwo{x,y\in\calS}{(x\ne y)}P_{(x,y)}\,\log\frac{P_{(x,y)}}{Q_{(x,y)}}\ge0,
\en
where the final inequality is the well known nonnegativity of relative entropy.~\qedm

\bigskip
It is worth noting that, when $q_x$ happens to be the canonical distribution, the right-hand side of \rlb{F:RRRR} reduces to the expressions \rlb{sigmai} or \rlb{sigmaii} of the entropy production rate.
\bigskip

We finally note that the method for proving \rlb{sigmaAA} leads to a general lower bound for relative entropy
\eq
D(p|q):=\sum_{x\in\calS}p_x\log\frac{p_x}{q_x}
=\sum_{x\in\calS}\Bigl(p_x\log\frac{p_x}{q_x}+q_x-p_x\Bigr)
\ge c_0\sum_{x\in\calS}\frac{(p_x-q_x)^2}{p_x+q_x}
=c_0\,\Delta(p|q),
\lb{Dlower}
\en
where $(p_x)_{x\in\calS}$ and $(q_x)_{x\in\calS}$ are arbitrary probability distributions, and the constant $c_0$ should satisfy $c_0\le8/9$ (or $c_0\le c^*=0.89612\cdots$).
We used \rlb{E:Hara2}.
The quantity $\Delta(p|q):=\sum_{x\in\calS}{(p_x-q_x)^2}/({p_x+q_x})$ is known as triangular discrimination.

As far as we know  the lower bound \rlb{Dlower} was derived by one of us (NS) and Takashi Hara, but we think it likely that it can be found in the literature.
In fact a slightly weaker version where $c_0$ is replace by $27/32$ appears, e.g., as (4.38) of Inder Jeet Taneja, {\em Bounds On Triangular Discrimination, Harmonic Mean and Symmetric Chi-square Divergences}\/({\tt arXiv:math/0505238v1.pdf}).  See also references therein.
We thank Sumio Watanabe for letting us know of the relevant references.

\end{widetext}


\begin{thebibliography}{99}
\bibitem{en:size}
More precisely, the size and the kinetic energy in this paragraph mean those of the parts of the engine that interact with the baths.
\bibitem{en:ratchet}
With a trivial extension, one can include a ``particle'' representing a macroscopic body (such as a ratchet) which forms apparatus constructing the engine.
\bibitem{en:td}
Extension to heat baths with time-dependent temperatures is trivial.
\end{thebibliography}
\end{document}